\documentclass[onecolumn,authoryear]{els-mrw} 

\usepackage{amsmath,amssymb,amsfonts,amsthm,makeidx,graphicx}
\usepackage{txfonts}
\usepackage{helvet}

\begin{document}

\chapter{Terrestrial planet surfaces and interiors}\label{chap1}

\author[1]{Ana-Catalina Plesa}%
\author[1]{Julia Maia}%
\author[1]{Solmaz Adeli}%
\author[1,2]{Tina Rückriemen-Bez}%

\address[1]{\orgname{German Aerospace Center}, \orgdiv{Institute of Planetary Research}, \orgaddress{Rutherford Str. 2, 12489 Berlin, Germany}}

\address[2]{\orgname{University of Münster}, \orgdiv{Institut für Planetologie}, \orgaddress{Wilhelm-Klemm-Str. 10, 48149 Münster, Germany}}

\articletag{Chapter Article tagline: update of previous edition,, reprint..}

\maketitle

\begin{glossary}[Nomenclature]
\begin{tabular}{@{}lp{34pc}@{}}
APL & Applied Physics Laboratoty\\
DAVINCI & Deep Atmosphere Venus Investigation of Noble gases, Chemistry, and Imaging\\
ESA & European Space Agency\\
FUB & Freie Universität Berlin\\
GRACE & Gravity Recovery and Climate Experiment\\
GRAIL & Gravity Recovery and Interior Laboratory\\
GSFC & Goddard Space Flight Center\\
InSight & Interior Exploration using Seismic Investigations, Geodesy and Heat Transport \\
JHU & John Hopkins University\\
JPL & Jet Propulsion Laboratory\\
KREEP & K (for potassium) REE (for rare earth elements) P (for phosporus)\\
MESSENGER & Mercury Surface, Space Environment, Geochemistry, and Ranging\\
MSSS & Malin Space Science Systems\\
NASA & National Aeronautics and Space Administration \\
PKT & Procellarum KREEP Terrane\\
SAR & Synthetic Aperture Radar\\
VERITAS & Venus Emissivity, Radio Science, InSAR, Topography, and Spectroscopy\\
\end{tabular}
\end{glossary}

\begin{glossary}[Glossary]
\begin{tabular}{@{}lp{34pc}@{}}
Adiabatic gradient & The change in temperature due to pressure without exchanging heat with the surroundings\\
Aeolian processes & Erosion, transport, and deposition processes driven by wind at or near the surface of rocky bodies\\
Alluvial fan & Landform that typically forms at foothills due to the accumulation of sediments transported by flowing water\\
Asthenosphere & Low-viscosity layer located in Earth's upper mantle, below the lithosphere\\
Brittle deformation& Type of deformation in which the material fractures when subject to external stress \\
Corona (coronae pl.) & Circular tectono-volcanic structure observed on the surface of Venus\\
Ductile deformation& Type of deformation in which the material flows and permanently distort when subject to external stress \\
Endogenic process & Geological process driven by the internal geodynamic activities of a planetary body\\
Eutectic mixture & A homogeneous mixture with a melting point lower than its constituents\\
Exogenic process & Geological process driven by forces external to a planetary body\\
Fluvial process & The natural process of physical interaction between the flowing water and the base rock\\
Granite & Coarse-grained igneous rock that forms from silica-rich magmas\\
Heat-pipe regime & Geodynamic/tectonic regime for which the main mode of heat transport is by volcanism \\
Impact gardening & The process through which impacts stir and mix the upper part of the crust of planetary bodies without an atmosphere\\
Lacustrine environment & Environment related to the presence of a body of water\\
Mantle plume & Upwelling of hot material transporting heat from the deeper interior towards the surface of a planetary body\\
Maria & Dark, smooth terrains on the Moon formed by basaltic lava flows \\
Moraine & Accumulation of unconsolidated material transported by glaciers\\
Plate tectonics regime & Geodynamic/tectonic regime for which the main mode of heat transport is by surface recycling\\
Plagioclase feldspar & Common rock-forming mineral composed of aluminum silicates combined with sodium and calcium\\
Plutonic squishy lid regime & Geodynamic/tectonic regime which is characterized by high rates of intrusive magmatism\\
Regolith & Loose, unconsolidated mixture of rock and dust on a planetary body's surface\\
Rheology & Mechanical property of a material that determines how it deforms under external stresses \\
Rift & Linear tectonic structure associated with extensional stresses\\
Shield volcano & A type of volcano with low-angle slopes resembling a shield lying on the ground\\
Space weathering & Changes that occurs in the chemical and physical structure of the surface material due to the material exposure to the sun and cosmos radiation and high energy particles\\
Stagnant lid regime & Geodynamic/tectonic regime where an immobile surface layer through which heat is transported only by conduction\\
Stratovolcano & A conical-shape volcano built by layers of volcanic material accumulated during successive eruptions\\
Subduction & The geologic process occurring when a tectonic plate slides beneath another plate and sinks into the mantle\\
Tessera & Highly tectonized terrains on Venus, typically associated with high elevation\\
Visible wavelength range & the part of the electromagnetic spectrum that is visible to human eye; 380-750 nanometer\\
\end{tabular}
\end{glossary}

\begin{abstract}[Abstract]
Rocky planets in our Solar System, namely Mercury, Venus, Earth, Mars, and the Moon, which is generally added to this group due to its geological complexity, possess a solid surface and share a common structure divided into major layers, namely a silicate crust, a silicate mantle, and an iron-rich core. However, while all terrestrial planets share a common structure, the thickness of their interior layers, their bulk chemical composition, and surface expressions of geological processes are often unique to each of them. In this chapter we provide an overview of the surfaces and interiors of rocky planets in the Solar System. We list some of the major discoveries in planetary exploration and discuss how they have helped to answer fundamental questions about planetary evolution while at the same time opening new avenues. For each of the major planetary layers, i.e., the surface, the crust and lithosphere, the mantle, and the core, we review key geological and geophysical processes that have shaped the planets that we observe today. Understanding the similarities and differences between the terrestrial planets in the Solar System will teach us about the diversity of evolutionary paths a planet could follow, helping us to better understand our own home, the Earth.
\end{abstract}

\section{Introduction}\label{chap:intro} 
\begin{figure}[h!]
\centering
\includegraphics[width=\textwidth]{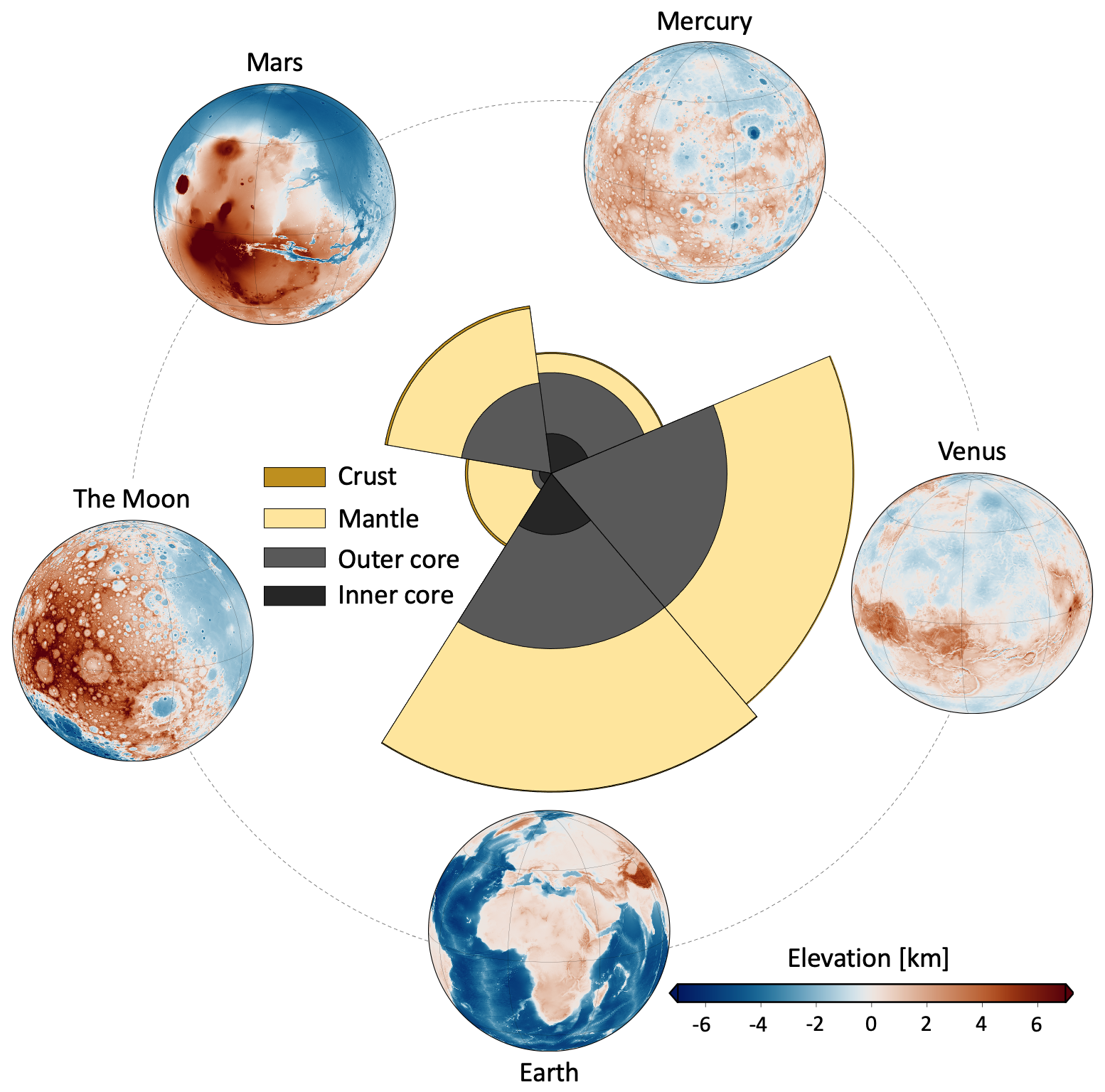}
\caption{Schematic view of the interior of terrestrial planets showing major chemical layers (i.e., the crust, the mantle, and the core subdivided into an outer liquid and an inner solid core). For each terrestrial planet the surface elevation is shown with the same color scale for comparison using orthographic projection.}
\label{Fig:Planets}
\end{figure}

The surfaces of terrestrial planets are windows into their interiors both in time and space, as they record tectonic activity caused by heating, cooling, and overall dynamics of the planets, volcanic features as witnesses of magmatic processes, as well as fingerprints of once active magnetic fields. All terrestrial planets share an interior structure made up by three major geochemical layers: the silicate crust, the rocky mantle, and the iron-rich core (Fig. \ref{Fig:Planets}). Some of these layers, such as the rocky mantle and iron-rich core have formed early during the initial planetary differentiation. Others, such as the crust and the presence or absence of a solid inner core, are witnesses of the long-term magmatic differentiation and planetary cooling, respectively. The thickness and composition of the crust are closely linked with mantle melting processes occurring deep in the interior, thus they provide crucial information to understand the magmatic differentiation of planets. The core - the innermost part of the planet - is fundamental to understanding the evolution of an intrinsic magnetic field. Its size in turn is critical to determine the thickness of the mantle, which is sandwiched between core and crust. The mantle is the main driver of the internal dynamics of planets, controlling their thermal history. Although all these layers (i.e., the surface, the crust, the mantle, and the core) are discussed separately in detail in the following sections, they are intimately connected, affecting each other's evolution. 

Earth is the planet where we have a vast amount of data at all spatial scales and ranging from the surface to the inner solid core located deep inside the planet, thanks to in situ measurements, laboratory analyses, and numerical models. However, Earth is perhaps the most complex planet among the terrestrial bodies, with its outer layer broken into tectonic plates, like pieces of a jigsaw puzzle, that are continuously moving on top of the underlying mantle, a characteristic not shared by the other terrestrial planets. Most of the information we possess about the other terrestrial planets today is by virtue of planetary missions and exploration that started back in 1959 with Luna 2, the first probe to reach the surface of the Moon, followed in 1962 by Mariner 2, the first successful interplanetary mission that performed a Venus flyby, proving that the planet has a surface temperature of 470°C. Seven years later, in 1969, another historic moment was recorded with the first crewed mission to successfully land on the surface of the Moon. Since then, a variety of planetary missions visited the planets and moons of our Solar System providing us with most valuable datasets, which, combined with knowledge about the Earth, can help us understand fundamental geological and geophysical processes. While a comprehensive overview of the missions flown to the terrestrial planets so far is beyond the scope of this chapter \citep[see][for an overview]{launius2013}, below we give a few examples about some of the major discoveries that helped us advance our understanding about the surfaces and interiors of terrestrial planets.

So far, only two missions made it to Mercury -- the first planet from the Sun with a diameter barely larger than our Moon. Mariner 10, launched in 1973, revealed an active magnetic field that is produced in Mercury's core. In fact, Mercury is the only terrestrial planet besides Earth that powers a core-dynamo today. Another surprising finding about Mercury are ice deposits at the poles, despite the proximity to the Sun. These ice deposits were first observed in 1990 as bright reflections in radar measurements from the Earth and later confirmed to be packs of frozen water by the MESSENGER mission in 2012 \citep[see ][for a comprehensive review on Mercury after MESSENGER]{solomon2018mercury}. Moreover, against expectations, MESSENGER found that Mercury's surface also holds other volatiles such as Potassium and Sulfur, and has evidence of explosive volcanism, a form of volcanism that requires the presence of volatiles in the subsurface \citep[][and reference therein]{solomon2018mercury}. How did Mercury retain its volatiles in such a harsh environment and why does it still have an active magnetic field are some of the major unanswered questions that future missions, such as BepiColombo currently en route (as of 2024) to the smallest of the terrestrial planets, will address.

Although Venus is one of our direct neighbors, its surface is challenging to observe from Earth, given the thick cloud deck that prevents the observation of the surface in the visible range. Hence, it was only with the development of radio science and  the onset of the space era that the surface of Venus was seen for the first time. Our view of the planet completely changed from a jungle world, an image painted by science fiction novels, to a world hostile to life with a crushing atmosphere and scorching hot surface temperature. The first and only global, high-resolution surface mapping of Venus was done in the 1990's by the Magellan spacecraft via Synthetic Aperture Radar (SAR) imaging. The mission showed that Venus' has only about 900 impact craters, indicating that the surface is overall young, a characteristic shared with the surface of the Earth \citep[see][for an overview of Venus science]{phillips1997venus}. Moreover, recent reanalysis of Magellan data have shown surface changes between images obtained eight months apart. These changes have been attributed to a volcanic eruption and indicate that Venus, similar to the Earth, is probably volcanically active today \citep{herrick2023}. Venus and Earth might have yet another common feature, namely continental crust which is inferred from spectral measurements of the Venus Express mission and from thermal evolution models. Future missions to Venus in the next decade such as VERITAS, EnVision, and DAVINCI, aim to provide long-awaited answers to open questions such as whether Venus had once similar surface conditions to the Earth, when in their history did the two planets start to diverge, and what is the level of tectonic and volcanic activity that Venus experiences today. 

The Moon is the first and only planetary body so far explored by crewed missions. Given its proximity to Earth, the Moon became an important target for early planetary exploration, with the Luna and Apollo programs. Our knowledge about the interior of the Moon relies to a great extent on the seismic and heat flow measurements performed by the Apollo missions. It is thanks to their measurements that the first models of the Moon's internal structure could be developed, showing an interior structure similar to the Earth with an iron core at the center, a silicate mantle, and a crust on top. Geochemical analyses of lunar samples collected by astronauts and returned to Earth led to the conclusion that the Moon experienced a massive melting event in its early history, forming a magma ocean. Later, global compositional and topography maps acquired by Lunar Prospector and Lunar Reconnaissance Orbiter, respectively, revealed that the nearside (i.e., the part of the Moon always facing the Earth) and the farside are different, with basaltic material mostly concentrated on the nearside, and the farside dominated by a different type of crust formed during the magma ocean solidification \citep[see][for the most recent and comprehensive review on the Moon]{Neal2024}. Modern "radio science" experiments performed by the GRAIL mission revealed the subsurface of the Moon with unprecedented detail, indicating the presence of magmatic intrusions within the lunar crust and providing crucial information about the magmatic history of the Moon \citep[][and references therein]{Neal2024}. Recently, the Moon came back into focus of international space agencies and commercial enterprises, with the first commercial Moon landing taking place in February 2024. As the Moon has much more to teach us about planetary formation and evolution, and with it being the perfect test ground for space technologies and in-situ resource utilization, the exploration of the Earth's natural satellite has just begun a new chapter.

Our view of the martian surface acquired by telescope observations in the late 19th and early 20th century, suggesting irrigation channels built by an intelligent civilization, has changed completely with the first flyby of Mariner 4 in 1965. Data returned by Mariner 4 and other missions showed a barren surface with no liquid water. Today, however, thanks to planetary exploration, we know that Mars once possessed liquid water at its surface that created some of the most spectacular features acquired by cameras on orbiting and landed missions. In 2005, a major breakthrough was recorded as for the first time hydrated minerals were detected at the martian surface by the Mars Express spacecraft, indicating that the surface experienced long periods with liquid water in the past, with conditions perhaps similar to the present-day Earth. While some of this water was lost to space, a substantial amount is believed to be locked in the martian subsurface in form of ice and groundwater. In fact, water ice has been identified in the ejecta material of the largest meteoroid impact ever witnessed on Mars that struck the martian surface on the Christmas Eve of 2021 and was detected by seismic measurements of the InSight lander. The resulting crater was observed by the Mars Reconnaissance Orbiter, making this the closest location to the equator where water ice has so far been found on Mars \citep{posiolova2022}. The InSight mission also provided the first seismic data from Mars using seismic waves to ``image'' its inner layers with unprecedented detail, much like ultrasound waves are used in medical imaging to look inside a human body. Although Mars is one of the most investigated planets beside Earth, many open questions remain, such as what is the rate at which Mars cools through time, what is the mechanism through which an early magnetic field, whose fingerprints are stored in the crust, was generated on Mars, how did Mars evolve to its current state, and did life ever develop on Mars, with the latter of these questions being one of the core objectives of the upcoming Mars Sample Return mission.

In the following sections we give a broad overview of general concepts describing the surfaces and interiors of terrestrial planets. Starting from the surface and going deeper into the interior, we describe the geological and geophysical processes that shaped the planetary surfaces that we observe today (Section \ref{chap:surfaces}), characterize the crust and lithosphere (Section \ref{chap:crusts}), drive the dynamics inside the rocky mantles (Section \ref{chap:mantles}), and affect the evolution of the iron-rich cores (Section \ref{chap:cores}). We then conclude by discussing how future planetary exploration beyond the currently planned missions could advance our knowledge about the surfaces and interiors of rocky planets in our Solar System.

\section{Surfaces}\label{chap:surfaces} 
The planetary surface refers to the solidified layer that forms the outermost portion of the crust and separates the planet’s interior from the atmosphere and/or outer space. The surface of a planet is then shaped by exogenic (Section \ref{sec:exogenic}) and endogenic processes (Section \ref{sec:endogenic}).  The best known and well-studied planetary surface is Earth's. Earth's surface is, however, heavily modified by biological processes and mostly covered by oceans. In addition, the majority of the Earth's oldest crust has been recycled by subduction due to plate tectonics, a process so far only known to Earth.

The surface of the inner Solar System planets is to some extent covered by a layer of unconsolidated and loose material called \textit{regolith}, the thickness and coverage of which varies depending on the planetary body. The lunar regolith is the most studied and best understood regolith thanks to the numerous in situ and remotely sensed measurements as well as returned samples by the Apollo missions \citep{Heiken1991}, as shown in Fig. \ref{Fig:surface}b. The regolith's physical and chemical characteristics of each planet is dependent and controlled by the endogenic, as well as by exogenic processes, such as space weathering, impact cratering and gardening. The regolith thickness, grain size, spectral response, and composition indicate its formation process and maturity, which in turn, inform us about the age of the surface, the composition of the bedrock, and processes involved. The material transfer over the surface is a powerful mechanism to shape landscapes. 

\subsection{Exogenic processes}\label{sec:exogenic}
It is critical to understand the mechanisms such as \textit{weathering and erosion} that transform the rock into the sediments and therefore, generate and modify the regolith. Chemical weathering refers to chemical changes in the material, such as change of mineralogy and composition; physical weathering deals with breaking down the fragments such as fracturing and swelling \citep[][Chapter 7]{Melosh2011}. Chemical weathering is a major topic when it comes to planets with an atmosphere and, most importantly, with liquids flowing over their surface. The best example is Mars with its liquid water-rich past which has attracted the major part of search for life investigation. Another process that can as well chemically and physically modify the surface of a planet, is known as \textit{space weathering}, which plays a major role on airless bodies, such as the Moon and Mercury. As stated by \cite{Pieters2016} space weathering is the gradual alteration of materials when they are exposed to a variety of natural processes that occur in the space environment. Irradiation by electromagnetic radiation or atomic particles from space and impacts from micrometeorites are the two main actors that cause space weathering. The returned samples from the Moon have greatly helped advance our knowledge in the chemical and physical modification through these processes.

\textit{Water and ice} are two major agents of forming landscapes. Fluvial and glacial processes and landscapes are common and familiar on Earth and on Mars, as detailed in \cite{Melosh2011}, Chapters 10 and 11. Understanding the source of the liquid and its cycle enables a better understanding of the local and global climate evolution. Water, in liquid form, carves valleys, channels, and riverbeds through erosion and forms alluvial and fluvial deltas and fans, as well as other depositional landforms through sediment transport, as exemplified in Fig. \ref{Fig:surface}c. Ice-driven processes occur when glaciers form, move, and erode. As glaciers move, they erode the underlying rock, carving U-shaped valleys, fjords, and creating various forms of moraines from the transported debris (Fig. \ref{Fig:surface}d). Water and ice can also weather rocks by seeping into cracks, freezing, expanding, and breaking them apart. They also alter the surface minerals to a new composition; a well-known example is the formation of the evaporate minerals at the surface of sediments in lacustrine environments. 

\textit{Mass wasting} or slope movement, is another process often observed on planetary surfaces. It is a process by which soil, rock, and debris move down a slope due to gravity \citep[][Chapter 8]{Melosh2011}. The result of such movement reveals great details about the composition, cohesion, and consistency of the surface material. Other mechanisms that affect the surface on planets with an atmosphere are \textit{aeolian} processes, driven by wind, eroding and transporting material on the surface. Features such as dunes (Fig. \ref{Fig:surface}e), yardangs, loess deposits, deflation hollow or blowout, and desert pavements are associated with aeolian mechanisms. 

\textit{Impact cratering}, illustrated in Fig. \ref{Fig:surface}f, is the most common feature to observe on a planetary surface. Studying the size-frequency distribution of craters is the best way of constraining the age of a planetary surface where radiometric dating is unavailable. In brief, a young surface has less time to be exposed to bombardment by impacts, and therefore possesses a low number of craters. Most of the impact craters on Earth (and Venus) have been erased by geological processes, but other inner Solar System planets have kept a record of the collision of asteroids, meteorites, and comets with their surface. Continuous mixing and disturbing the upper part of the planetary surface is called impact gardening.

\subsection{Endogenic processes}\label{sec:endogenic}
\begin{figure}[h!]
\centering
\includegraphics[width=0.9\textwidth]{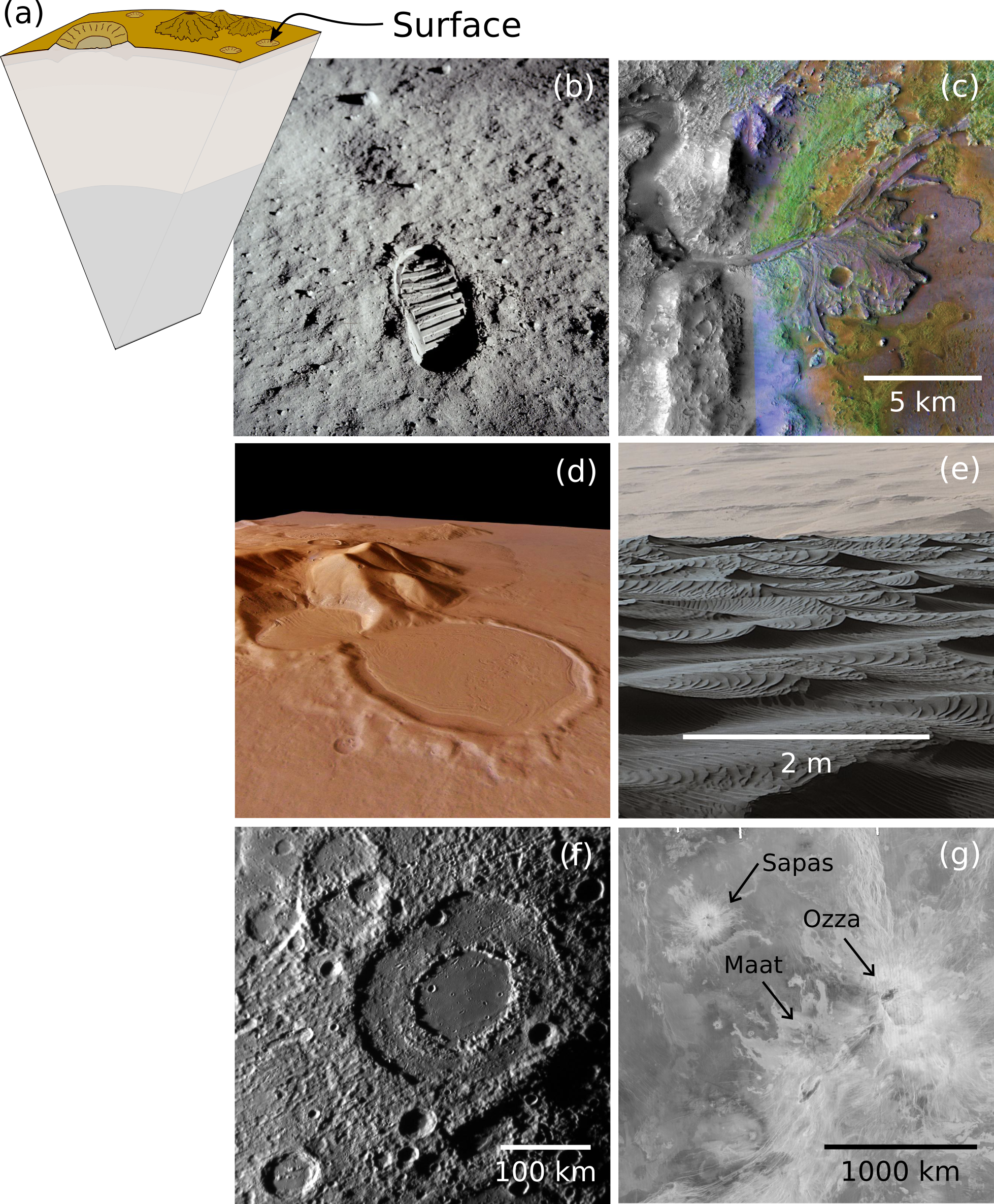}
\caption{(a) Schematic view of a planetary surface. (b)  Lunar regolith and the Buzz Aldrin footprint during the Apollo 11 mission (NASA). (c) River delta on the Jezero Crater on Mars. This figure, which combines images obtained by the context camera and spectral information from CRISM, both instruments onboard of the Mars Reconnaissance Orbiter, show clays and carbonate sediments which indicate chemical alteration by water (NASA/JPL-Caltech/MSSS/JHU-APL). (d) Perspective view of two 'hourglass' shaped craters on Mars which correspond to a glacial deposit flowing from the flank of a hill into a small crater  (9\,km of diameter) followed by a flow into the large crater (ESA/DLR/FUB). (e) Section of the Bagnold Dunes field on Mars observed by the Mast Camera onboard of the Curiosity Rover (NASA/JPL-Caltech/MSSS). (f) Peak-ring Michelangelo impact basin on Mercury observed by MESSENGER's MDIS camera (NASA/JHU-APL). (g) Volcanoes and rifts at Atla Region on Venus observed by the  Synthetic Aperture Radar on board on the NASA Magellan spacecraft. Arrows and labels indicate the name and location of major shield volcanoes in the region (NASA/JPL). Note that on radar image in panel (g), bright colors generally represent rugged terrains and dark colors indicate smooth regions.}
\label{Fig:surface}
\end{figure}

Processes occurring inside a planetary body also play an important role in shaping structures and features on planetary surfaces. Venus is an example of a planet where the surface has mainly been shaped by endogenic processes (Fig. \ref{Fig:surface}g). The most common processes are \textit{tectonism} and \textit{volcanism}. Planetary surface features formed by internal stresses that crack or deform the lithosphere are collectively known as tectonic features \citep[][Chapter 4]{Melosh2011}. Tectonic activity is caused by internal stresses driven by heat loss; all the terrestrial planets are in a cooling phase, after having passed through a molten (or nearly molten) stage early in their development. Analyzing the faults, folds, scarps, ridges, and troughs formed by tectonic processes provides insight into the forces that have affected a planet's lithosphere. By studying the topography and understanding the surface material's rheology, it is possible to infer the direction and magnitude of stresses.

\textit{Volcanism} is the process describing the eruption of molten rock, or magma, from beneath the crust to the surface, forming volcanic landforms \citep[see][Chapter 5, for details]{Melosh2011}. The internal heat of a planet drives this processes, initiated by melting of rocks in a planet’s interior, followed by magma rising to the surface through cracks or vents. There are different types of volcanic landforms on Earth such as shield volcanoes, stratovolcanoes, and lava plains that are characterized by different magma viscosities and compositions, amount of gas in the magma, or magma migration processes. An example of a shield volcano, typically characterized by low viscosity magma, is Olympus Mons on Mars, the highest volcano in the Solar System. Lava plains or lava fields, which are large, flat areas covered by highly fluid basaltic lava flows, are widely present on the inner Solar System planets. Stratovolcanoes, which form by high viscosity magmas and are likely associated with explosive eruptions due to the presence of gas in the ascending magmas, might have existed or might exist on other planets than Earth, e.g. Venus, but there is no clear evidence for it, yet.

\subsection{The planets}
\subsubsection*{Mercury}
The heavily cratered surface of Mercury is very similar to the Earth's Moon, and it appears gray-brown in the visible wavelength range. Mercury features some extreme environments: being the closest planet to the Sun, the surface temperature can reach about 450°C. However, due to the complete lack of atmosphere, the night temperature is as cold as -170°C. This temperature swing is unique in the Solar System. In addition, the planet is exposed to extreme high solar radiation, making space weathering a major process that modifies its surface. 

Mercury is currently geological inactive, and its lack of atmosphere means that impacts from asteroids and comets have shaped its surface for billions of years and are well preserved with little erosion to modify them. Large impact basins are prominent, such as Caloris Basin (1,550\,km in diameter), one of the largest in the Solar System, along with large areas of smooth terrains, and cliffs, formed as the planet's interior cooled and contracted over  billions of years. Mercury is rich in peak rings, which are large impact craters characterized by a ring of peaks surrounding the center of the crater, rather than a single peak (Fig. \ref{Fig:surface}f).

Mercury's surface has a surprising lack of iron, which is unusual for a planet with a massive iron core. Additionally, lighter elements like sulfur and potassium are more abundant, suggesting that Mercury’s original crust and mantle materials may have been eroded or buried. More investigations are crucial in order to answer these questions. This is one of the main objectives of the BepiColombo mission, with an arrival date of November 2026.

\subsubsection*{Venus}
Although Venus is very similar to Earth, in term of size and density, its surface conditions are much more hostile: with a surface pressure of $\sim$92 bars and temperature of more than 450\textdegree C, Venus is the hottest planet in the Solar System. Moreover, Venus is globally covered by thick sulfuric acid clouds blocking the visible and near-infrared light, but radar imaging has revealed a large variety of geological features. Its surface shows wide-spread volcanism, with vast volcanic plains, thousands of shield volcanoes, and numerous lava flows. Venus also has unique tectonic structures, such as the tesserae, which correspond to terrains with multiple stages of deformation, and coronae, approximately circular volcano-tectonic structures, whose origin is typically associated with mantle upwellings. In addition, the planet shows compressive ridge structures and long rift zones caused by extensional tectonics. Venus' surface has very few impact craters, revealing a young surface with an age of less than 750\,Myr. The young surface combined with the diversity of structures indicate that Venus is geologically active, probably experiencing vigorous mantle convection at present day (Section \ref{chap:mantles}). 

Little is known about the surface composition and level of volcanic activity that might have occurred or still be happening on Venus. The only in situ data are from the Venera and Vega programs of the Soviet Union with several landers surviving for a short period of time on the surface. The future orbital missions EnVision and VERITAS are designed to map the rock type on the surface of this planet, and to investigate its topography and atmosphere in more detail, with the major goal to shed light on its past and current surface conditions and geological history.

\subsubsection*{The Moon}
The Moon is a small and airless body, similar to Mercury. Thus, the geological processes are mainly limited to external sources, such as impact of asteroids and space weathering. The lack of atmosphere allows a steady bombardment by minor bodies, turning the surface of the Moon into a blanket of rock fragments, boulders and regolith. The Moon's surface is divided to dark and light regions, called highlands and mare, respectively, which are composed of material with different composition, indicating the past geological history of the Moon and the early phases of crust formation (see Section \ref{chap:crusts}). 

The dayside temperature can reach arround 130\textdegree C while the nightside temperature can be as cold as -170\textdegree C. Despite the lack of an atmosphere and the high temperature on the dayside, there are certain areas on the Moon where water can exist in the form of ice. These are on the floor of impact craters located very close to the poles and have not received direct sunlight for millions or billions of years, called \textit{permanently shadowed regions}. Such regions have also been detected on Mercury, and are of high interest when it comes to search for water and life in the Solar System as well as to provide resources for future crewed missions to the Moon.

\subsubsection*{Mars}
Today, Mars is a hyperarid and frozen desert, but there is clear evidence of the presence of liquid water in its geological past. Its regolith is rich in iron oxide, giving its distinctive reddish color. One of the most prominent features on Mars is the dichotomy between the heavily cratered (older) southern highlands and the sparsely cratered (younger) northern lowland. The differences in altitude, that can reach over 4\,km between the two hemispheres, and the crater record show that the northern hemisphere is a large basin, which, at the early impact cratering phases of Mars, might have been either covered by ice and therefore protected from impacts or has been through a major resurfacing event. 

Common features on Mars are vast plains, deep canyons like Valles Marineris, and towering volcanoes such as Olympus Mons, the tallest in the Solar System. Mars also shows complex sedimentary processes. The thin carbon dioxide atmosphere, lacks sufficient pressure to support liquid water today, but surface observations indicate that Mars once had rivers and lakes. This evidence is geomorphologic: dry valley networks, system of lakes, alluvial and fluvial fans and deltas; and chemical: thick deposits rich in various clay minerals, sulfates, chloride-salts, and carbonates (Fig. \ref{Fig:surface}c). These minerals only form over geologically long periods of time during which liquid water and bedrock can interact. We now know that the red planet could have sustained life, should life have emerged on Mars. This has triggered numerous studies and missions to search for traces of extinct and/or extant life. Mars also features current and past signatures of glacial activity (Fig. \ref{Fig:surface}d), being the only planet besides the Earth to have ice polar caps. Similar to the polar caps on the Earth, the martian polar caps are made mostly of pure water ice, but include also contributions from CO$_2$ ice and dust.

\paragraph{}
\section{Crusts and lithospheres}\label{chap:crusts} 
\begin{figure}[h!]
\centering
\includegraphics[width=1\textwidth]{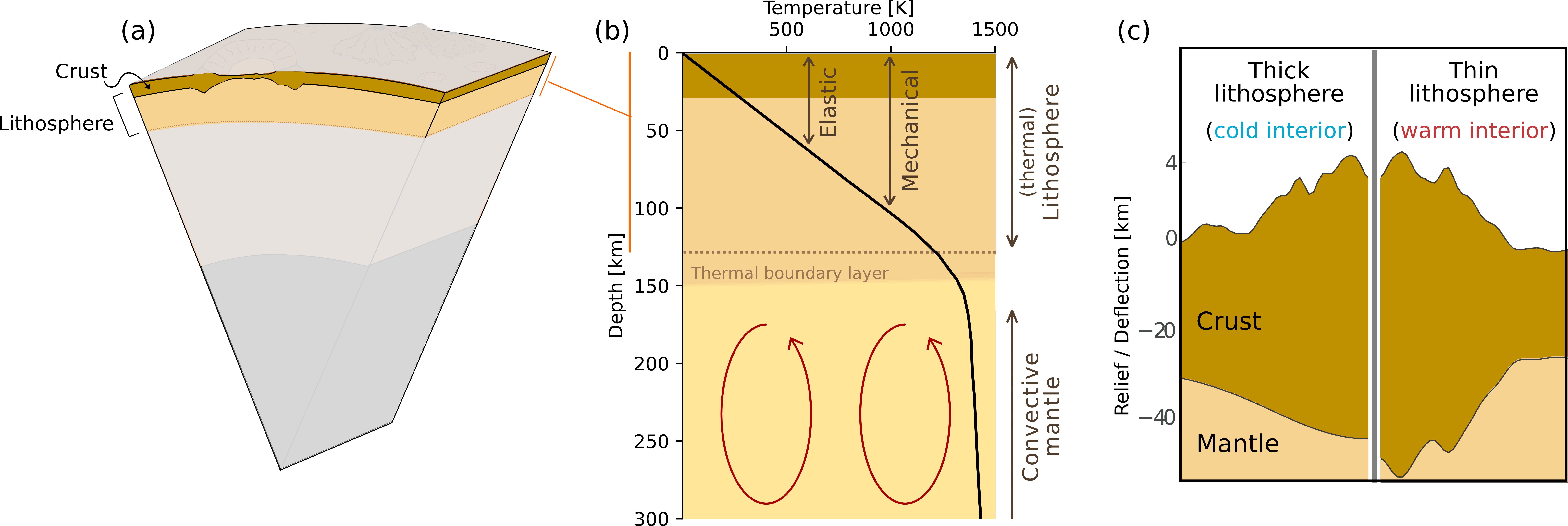}
\caption{Schematic view of the definition of crust and lithosphere. (a) Location of crust and lithosphere with respect to other layers. (b) Characteristic temperature profile associated with the upper mantle and its relation to different types of lithospheric layers, \textit{elastic} lithosphere, \textit{mechanical} lithosphere, and \textit{thermal} lithosphere. (c) Illustration of crustal thickness variations associated with topographic features. The lithosphere thickness affects the formation of crustal roots, since thicker (and stronger) lithospheres are able to support a larger weight of the topographic load in comparison to thinner (and weaker) lithospheres.}
\label{Fig:lit-scheme}
\end{figure}

Crusts and lithospheres -- the thin outermost layers of terrestrial planets -- are of uttermost importance for the understanding the formation and geological evolution of planets. The crust corresponds to a chemically differentiated layer mostly composed of rocks that are lighter, and therefore more buoyant, than the underlying mantle. The composition and thickness of the crust holds fundamental information about the differentiation and subsequent geological evolution of the planet as a whole. Meanwhile, the lithosphere is defined as the strong, outermost mechanical layer of planets which typically comprises the crust and the uppermost part of the mantle, being intrinsically related to the thermal state and interior evolution of planets.  Fig. \ref{Fig:lit-scheme} illustrates the key concepts of crust and lithosphere which are described in detail throughout this section.

\subsection{Crust: the outermost chemical layer}\label{sec:crust}
Although crusts are ubiquitous to rocky planets, their geophysical and petrological properties vary considerably due to different formation processes \citep[see][for details]{taylor2008}. To accommodate this variability while facilitating comparative analyses and discussions, a classification scheme that categorizes crusts as \textit{primary}, \textit{secondary}, or \textit{tertiary} is commonly used. 

\textit{Primary crusts} are the result of the initial magma ocean solidification, which happens in the first hundred million of years of the planet's evolution. After the accretion period, rocky planets - at this point almost entirely molten - differentiate into a metallic high-density core and a magma ocean. As the planet cools, the magma ocean starts crystallizing from the deep mantle towards the surface, a relatively fast process that takes about 100\,Myr. During the magma ocean solidification, magnesium-rich cumulates sink to form the mantle, while light anorthositic rocks float to the surface and form the primary crust.

Once the magma ocean crystallizes, the onset of global, solid mantle convection controls the planet's heat transport and triggers partial melt of silicate mantle rocks, typically in hot mantle upwellings (Section \ref{chap:mantles}). This melt is extracted from the mantle to the surface, generally forming basaltic rocks, or remains trapped in the existing crust and lithosphere, forming magmatic intrusions. This is the standard formation process of \textit{secondary crusts}, which are produced as long as the planet is warm enough to undergo partial melting of the mantle, typically associated with time spans of billions of years.

Geological processes can lead to the recycling of secondary crustal materials. These processes are commonly related to remelting of secondary crusts often associated with crustal recycling in the presence of volatiles that promote partial melting by lowering the melting temperature. These melts form what it is known as \textit{tertiary crusts}, characterized by silica-rich rocks, such as granites. The formation of tertiary crusts is inefficient in comparison to the other processes. The presence of such crust is rare and only proved to exist in large amounts on Earth, where it is the main constituent of the continental crust.

Along with the composition, the thickness of the crust is also fundamental to understand the geological history of a planet. The crustal thickness informs about the rate at which the crust has been produced and potentially recycled. Planets that have no effective mechanism for crustal recycling tend to have a relatively thick crust. Meanwhile, the internal temperature of planets, which is related to their size, will strongly influence the amount of melt and, consequently, how much secondary crust is produced. 

\subsection{Lithosphere: the outermost rigid layer}\label{sec:lithosphere}
In the earliest days of the field of geophysics the scientific debate was centered on the possible mechanisms responsible for the support of topographic features, which correspond to non-hydrostatic deviations from the hydrostatic shapes of planets. This debate led to the development of models of isostatic compensation, in which topographic loads are supported buoyantly as the crustal loads float on top of a denser yet weaker (i.e., viscous) mantle. In the early 20th century, scientists noticed that isostatic models were not sufficient to explain the local variations in the gravity field in large mountain ranges. In particular, these observations indicated that the Earth should have an uppermost layer with finite strength capable of partially sustaining the planet's topography. This rigid layer, which comprises the crust and the uppermost mantle, is the so-called lithosphere \citep{barrell1914}. Since then, the properties of the lithosphere have been extensively investigated with different sets of observations, models, and laboratory experiments, leading to a significant evolution of its meaning depending on the context it is discussed. As a result, different ``types" of lithosphere have been defined, as illustrated in Fig. \ref{Fig:lit-scheme}b. A detailed review on the structure and dynamics of the lithosphere can be found in \cite{watts2009treatise}.

Nowadays, when using the term ``lithosphere'' researchers are typically referring to the rigid, cold layer of a terrestrial planet in which  heat is mainly transported via conduction, as opposed to the underlying hot, convective mantle (see Section \ref{chap:mantles} for details). Being intrinsically related to heat transfer, this layer is also commonly called \textit{thermal lithosphere}. On Earth, the thermal lithosphere is divided into plates that move and can get recycled via subduction processes, representing the main actors of plate tectonics. The other terrestrial planets  (maybe with the exception of Venus) have an immobile, single lithospheric plate, known as the stagnant-lid regime.

A second important definition is the so-called \textit{mechanical lithosphere}, which corresponds to the rigid layer of terrestrial planets that can maintain high differential stresses typically caused by tectonic activity. The rheological properties of the mechanical lithosphere can be quite complex, but they can be usually subdivided into three layers: (1) An uppermost brittle zone where temperatures and pressures are low and rock deformation is associated with fracturing and faulting. (2) An elastic layer where deformation is a linear function of strain. (3) A ductile layer located at large depths where temperatures and pressures are high. Because in the ductile regime strength decreases exponentially with temperature, the rheology quickly transitions from a strong, highly-viscous layer, which correspond to the lowermost part of the mechanical lithosphere, to a deeper zone with a fluid-like behavior. 

One last type of lithosphere that merits discussion is the \textit{elastic lithosphere}. The emplacement of a load on top of the crust, such as a volcano, will bend the lithosphere downward. The response of the lithosphere to loading is controlled, to first order, by elastic deformation. Several studies have shown that employing a thin elastic shell model is a simple but accurate way to analyze the lithosphere response to loading \citep[see][for an overview]{watts2009treatise}. The thickness of this elastic shell, often called \textit{elastic thickness}, determines the amount of lithospheric flexure due to loading (Fig. \ref{Fig:lit-scheme}c). Thin elastic lithospheres are associated with low rigidity and large displacements, while thick layers lead to small displacements. These processes directly impact geophysical observables, such as the gravity anomalies and topography signatures. Hence, elastic thickness analysis is a powerful tool to study the lithospheres of terrestrial planets.

Although related to different properties, these three definitions of lithosphere are correlated. The elastic lithosphere is thinner than the mechanical lithosphere which, in turn, is thinner than the thermal lithosphere. Moreover, the thickness of these layers are strongly related to temperature and the overall thermal state of the planet. Terrestrial planets with cold interiors, typically the small ones, are associated with thick and strong lithospheres (elastic, mechanical, and thermal). Meanwhile, large planets, such as Earth and Venus, have hot interiors and relatively thin lithospheres. 

\subsection{The planets} \label{sec:crust-planets}
Each terrestrial planet has its own unique crustal and lithospheric properties, which unveil fundamental information about their geological and thermal evolutions. Fig. \ref{Fig:Tc-maps} summarizes some of the key characteristics of the planetary crusts.

\subsubsection*{Mercury}
\begin{figure}[h!]
\centering
\includegraphics[width=0.9\textwidth]{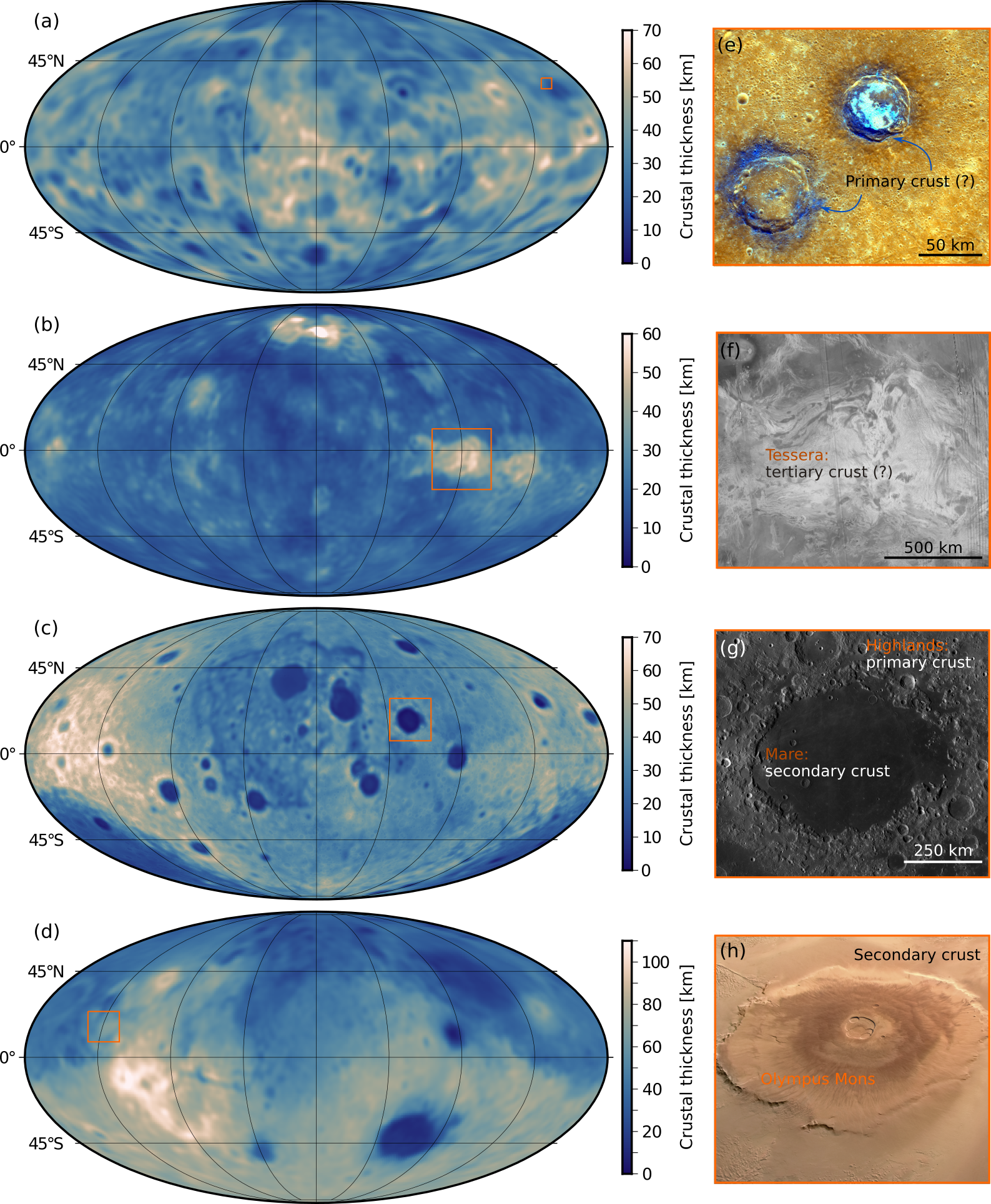}
\caption{Properties of planetary crusts, including crustal thickness maps (a-d) and surface images (e-h). Crustal thickness maps of (a) Mercury, (b) Venus, (c) Moon, and (d) Mars based on gravity and topography data analysis shown in Mollweide projection centered at 0°W longitude. More information on the geophysical datasets available for each planet and the techniques used here to estimate crustal thickness can be found in \cite{spohn2015}, Chapter 5. (e) Enhanced color image of Mercury seen by MESSENGER showing dark, blueish material linked to impact craters which might be associated with excavated carbon-rich primary crust (NASA/JHU-APL/Carnegie Institution of Washington). (f) Tessera Plateau on Venus observed by Magellan, hypothesized to be associated with silica-rich crust (NASA/JPL). (g) Mare Crisium on the Moon seen by the Lunar Reconnaissance Orbiter, illustrating the difference between dark, young mare lava flows versus bright, old highland terrains (NASA/GSFC/Arizona State University). (h) Perspective view of Olympus Mons on Mars, the tallest volcano in the solar system, reaching over 20\,km in altitude and 600\,km in diameter (ESA/DLR/FUB/AndreaLuck).}
\label{Fig:Tc-maps}
\end{figure}

Using gravity and topography data obtained by the MESSENGER mission, geophysical studies estimated that the average crustal thickness of Mercury should be between 15 and 50\,km. Considering that the core-mantle boundary is $\sim400$\,km deep, the crust of Mercury represents 5 to 15\% of the planet's silicate volume. This high fraction, potentially the largest one among the terrestrial planets, is consistent with a stagnant-lid planet and indicates that Mercury had a high efficiency of crustal production throughout its evolution. In addition, compositional and geological analysis of Mercury's surface show that the crust is mostly composed of secondary volcanic processes, although a relatively high amount of carbon on the surface ($\sim1$\% concentration) typically associated with regions excavated by impact craters (Fig. \ref{Fig:Tc-maps}e) suggest the presence of remnants of a primary crust generated by graphite flotation \citep[e.g.,][Chapter 3]{solomon2018mercury}.

Mercury has major lateral crustal thickness variations (Fig. \ref{Fig:Tc-maps}a). Regions of thick crust are typically located at high topography regions at low latitudes. Thin crust regions are associated with large impact basins and the so-called smooth plains near the north pole. The smooth plains are associated with large-scale effusive volcanism, which correspond to the youngest terrain on the planet, formed about 3.5 billions of years ago. This long-standing geological quiescence is in-line with thermal evolution models which indicate that the entire mantle of Mercury could be in a cold, conductive state at present-day, hence, currently, Mercury's thermal lithosphere might correspond to the entire silicate portion of the planet \citep[][Chapter 19 and references therein]{solomon2018mercury}.

\subsubsection*{Venus}
In the 1980's a fleet of Soviet landers provided the only direct major-element compositional constraints of Venus' surface to date. These data showed that the planet's surface is composed of basalts, similar to the ones formed on mid-ocean ridges on Earth. These constraints, combined with the tens of thousands of volcanic features observed on the surface (Section \ref{chap:surfaces}) indicate that Venus' crust is predominantly of secondary origin. Moreover, gravity and topography investigations, based on data obtained by the Magellan spacecraft, suggest that the crust is, on average, 20\,km thick, which represents $\sim$1\% of the planet's silicate volume \citep[see][and references therein]{rolf2022}. This volume is similar to the crustal volume on Earth. In addition, the surface of Venus is young (less than 750\,Myr old). These properties indicate that Venus probably experiences crustal recycling -- although the nature of this process is not well-understood since plate tectonics is not observed on Venus at present-day. Recently, it has been proposed that Venus could be experiencing regional recycling, associated with high amounts of intrusive magmatism which would weaken the crust and promote surface mobilization (see Section \ref{chap:mantles}). This unique tectonic regime, named `plutonic squishy-lid' is also consistent with the $\sim$100\,km thick thermal lithosphere estimated for Venus today \citep[][and references therein]{rolf2022, Gillmann2024}.

Lateral crustal thickness variations are overall mild in comparison to the other planets. Regions of thick crust are located in highlands near the equator and the north pole. The thick crust along with the low surface emissivity of these features observed by the Venus Express mission indicate that these regions could be analogous to continents on Earth and represent large volumes of tertiary crust - but this interpretation still has to be confirmed by future observations of the upcoming missions VERITAS and EnVision \citep[e.g.,][]{rolf2022}.

\subsubsection*{The Moon}
The Moon is probably the best object in the Solar System to study the contrast between primary and secondary crusts. About 17\% of the lunar surface is covered by dark basaltic lava flows, the so-called Maria. These basalts were mostly formed between 3.8 and 3.1\,Gyr ago but flows can be as young as 1\,Gyr old. Interestingly, the Maria are concentrated in a relatively small region on the lunar nearside. The cause of this spatial concentration is not fully understood, but it seems to be related to the so-called Procellarum KREEP Terrain (PKT), a region highly enriched in radioactive elements, that leads to regionally high lithospheric temperatures and facilitates partial melting and volcanism. Although wide-spread within the PKT, the mare basalts are on average about 1 to 2\,km thick, representing only a minor contribution to the total crustal volume. In fact, most of the lunar crust is composed of a bright, plagioclase feldspar-rich rock called anorthosite. The anorthositic crust is much older than the basalts, being formed roughly 4.4\,Gyr ago. The composition, global distribution, and ancient nature of these rocks led to the well-accepted hypothesis that most of the lunar crust formed via the early differentiation and solidification of a magma ocean. More details on the evolution of the lunar crust can be found in \cite{Neal2024}, Chapter 7.

Along with the compositional contrast, the near- and far-side of the Moon also present a dichotomy in terms of topography and crustal thickness. The far side is associated with higher topography and thick crust, being commonly called the lunar highlands, while the nearside, dominated by the mare basalts, has significantly thinner crust, as shown in Fig. \ref{Fig:Tc-maps}c. Studies combining  measurements from Apollo seismometers and gravity field data by the GRAIL mission indicate that the Moon has an average crustal thickness of 38\,km \cite[see][Chapter 5 and references therein]{spohn2015}, corresponding to $\sim6$\% of the silicate volume which is in line with the volume of other stagnant-lid planets. Thermal evolution models suggest a thick lunar lithosphere \citep[e.g.,][Chapter 1]{spohn2015}, which is typical of small and cold bodies.

\subsubsection*{Mars}

The latest crustal thickness estimates for Mars, which combine gravity and topography analysis with constraints from seismic data by the InSight lander, show that the the planet has a crust of about $50\pm 20$\,km on average \citep{wieczorek2022}. This value corresponds to $\sim 6$\% of the volume of the silicate portion of planet. Also, similar to the Moon and Mercury, Mars has a thick lithosphere of about 500\,km \citep[][and references therein]{breuer2022}.

Mars is the planet with the largest variability in crustal thickness. The Tharsis rise, a volcanic province that covers about 25\% of the martian surface, likely represents the region with thickest crust in the entire Solar System, reaching over 100\,km. In addition, Mars presents a north-south dichotomy, where the south hemisphere shows highlands, typically associated with a thick crust and surface ages of about 4\,Gyr, while the north hemisphere corresponds to a thinner crust with a smooth surface and ages younger than 3.7\,Gyr. Despite these differences, both hemispheres appear to have similar basaltic compositions, indicating that overall the martian crust has a volcanic (secondary) origin.  

\section{Mantles}\label{chap:mantles}
The mantle, located between the crust and core, influences the thermal history and interior dynamics of rocky planets. Its dynamics and thermal evolution depend on the interior structure, specifically the mantle thickness that in turn is determined by the core size, and crustal thickness. Radioactive heat acquired from the building blocks during the planet's formation provides one of the most important sources of heat in the mantle and affects the thermal state and cooling history of the interior. Additionally, mantle melting and differentiation processes throughout the evolution, as well as solid-solid phase transitions that occur when a different crystal structure becomes more stable than another can affect the dynamics and heat transport in a planet's interior.

\begin{figure}[h!]
\centering
\includegraphics[width=\textwidth]{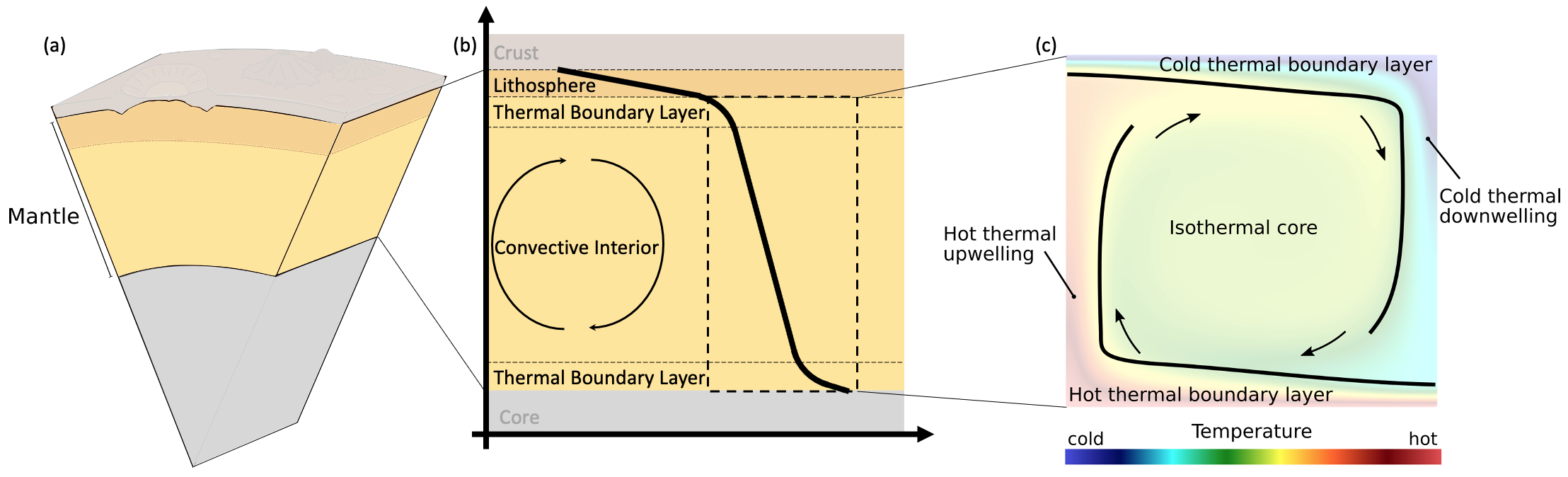}
 \caption{Schematic view of the silicate mantle layer. (a) Location of the mantle layer with respect to other major layers. (b) Typical temperature profile throughout the mantle including a thermal lithosphere, thermal boundary layers and a convecting interior. (c) The characteristics of a mantle convection cell modified after \citet{Schubert2001}.}
\label{Fig:Tprofile}
\end{figure}

In the following sections, we describe the heat transport (Section \ref{sec:convection}) and geodynamical regimes proposed for rocky planets in our Solar System (Section \ref{sec:geodynamic_regimes}). Section \ref{sec:mantle_planets} focuses on the dynamics of the mantles of Mercury, Venus, the Moon, and Mars. While not focusing on Earth, we describe the physical processes and the interior dynamics of these planets in comparison to the Earth.

\subsection{Mantle Convection}\label{sec:convection}
The mantle is typically subdivided into the thermal lithosphere, where heat is transported by conduction, and a deeper interior, where the heat transport is dominated by convection. Mantle convection is the slow creeping flow of silicate material that occurs on time-scales of millions and billions of years \citep{Schubert2001}. Convection is primarily driven by instabilities that occur in the thermal boundary layers, resulting in hot upwellings from the bottom thermal boundary layer and cold downwellings from the top thermal boundary layer (Fig. \ref{Fig:Tprofile}).

In 1862, William Thompson (Lord Kelvin) estimated the Earth's age based on its conductive cooling from a molten state, suggesting it was only a few tens to a few hundred million years old -- much shorter than contemporary geological estimates. A few years later, his former assistant John Perry pointed out that convection in the Earth's interior would change Kelvin's estimate, but his argument was disregarded at that time and later forgotten \citep{England2007}. While radioactivity, discovered later by Pierre Curie and Albert Laborde, was neglected as a source of heat in Kelvin's calculations and is often mentioned as the reason for Kelvin's mistake, it is by incorporating convection that the Earth's age aligns more closely with radiometric dating results \citep{poirier2017}. 

The vigor of convection in rocky planets is characterized by the \textit{Rayleigh number}, named after Lord Rayleigh, a British mathematician and physicist who developed the mathematical theory of fluid dynamics. This non-dimensional number compares factors driving convection (like temperature difference across the mantle and mantle thickness) to those resisting it (such as viscosity). While some parameters, like viscosity and temperature contrast, have significant uncertainties, the range of Rayleigh numbers of terrestrial planets indicates that Earth and Venus likely have vigorous convection, while Mercury's thin mantle may have cooled and reached a conductive state at present day.

\subsection{Geodynamic Regimes}\label{sec:geodynamic_regimes}
The dynamics and cooling of the mantle of terrestrial planets are influenced by their geodynamic regimes, which affect surface tectonics, crust and lithosphere thickness, and core cooling. Several major regimes have been proposed to describe the heat transport and interior dynamics \citep{Lourencco2023}. In the \textit{plate tectonics regime}, convection involves the outermost layer, where cold surface material is subducted and hot mantle is exposed, resulting in a young surface and efficient cooling. This is exemplified by present-day Earth, but a similar geodynamic regime has been proposed for Venus and Jupiter’s moon Europa. However, the early Earth likely had a different regime, such as the heat-pipe or the plutonic squishy lid regime, where heat transport was dominated either by extrusive magmatism or by intrusive melt trapped in the lithosphere, respectively. The \textit{heat-pipe regime} features a thick, cold lithosphere that traps heat and prevents it from efficiently escaping the deep interior. While this regime has been proposed for the early Earth, it best describes Jupiter's moon Io, the most volcanically active rocky body in our Solar System. In contrast, the \textit{plutonic squishy lid regime} is characterized by a hot, thin lithosphere that can mobilize due to magmatic intrusions, allowing for surface recycling on a smaller scale than plate tectonics. This regime is thought to characterize both early Earth and Venus. Smaller terrestrial planets like Mars, Mercury, and the Moon, however, are considered to operate under the \textit{stagnant lid regime} (so-called one plate planets), with their billions-of-years-old surfaces showing no signs of recent surface mobilization or intense magmatic activity.

\subsection{The planets}\label{sec:mantle_planets}
\begin{figure}[h!]
\centering
\includegraphics[width=\textwidth]{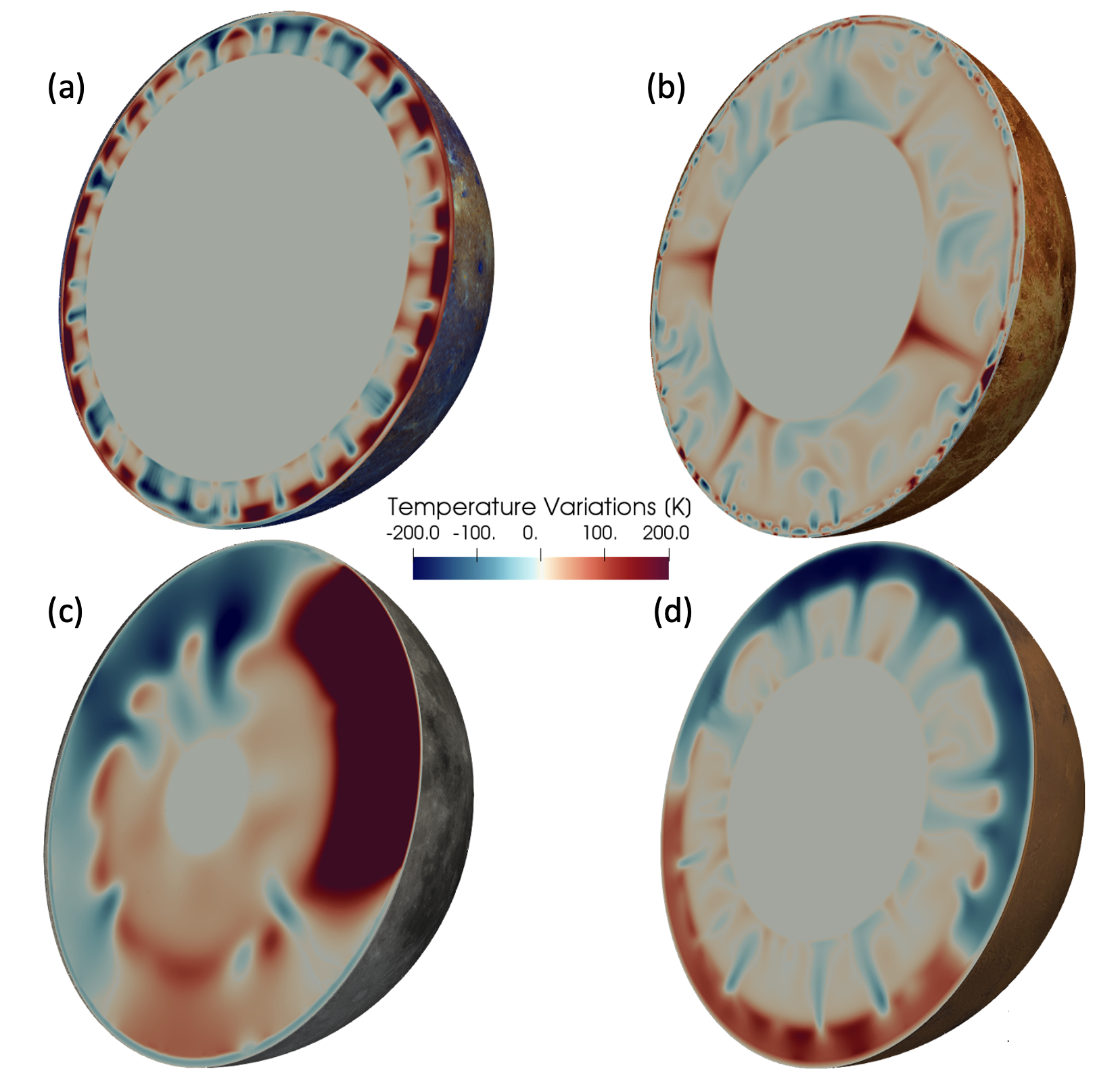}
\caption{Convection pattern obtained when using (a) Mercury-like parameters, (b) Venus-like parameters, (c) Moon-like parameters, and (d) Mars-like parameters. The snapshots illustrate temperature variations during the planetary evolution. Temperatures that are warmer than the average are shown in red, while regions colder than the average are illustrated in blue.}
\label{Fig:Convection}
\end{figure}

The different interior structures, surface temperatures, and crustal thickness variations of the terrestrial planets affect the near-surface thermal state and the dynamics in the deep interior. Fig. \ref{Fig:Convection} illustrates the convection pattern obtained for Mercury-, Venus-, Moon-, and Mars-like parameters. 

\subsubsection*{Mercury}
Mercury, the smallest terrestrial planet, has an extreme interior structure, with its core making up about 85\% of its radius—compared to Earth's core, which is about 55\% of its total radius. This results in a thin silicate mantle of just 400\,km, and consequently a small silicate volume. Mercury possesses the smallest silicate volume to surface area ratio among the terrestrial planets, which likely led to substantially cooling during its history, with major consequences for the thermal history and present-day thermal state. During the evolution, compressional tectonic features (i.e., wrinkle ridges, lobate scarps) formed at the surface of Mercury, similar to aging fruits that become wrinkled with time. These tectonic features have been used to estimate the radius decrease (i.e., contraction) due to planetary cooling, likely suggesting that Mercury possesses the most pronounced planetary contraction among the terrestrial planets \citep[][Chapter 10 and references therein]{solomon2018mercury}. While Mercury's mantle was once convective, it may now be entirely conductive due to this significant cooling.

Mercury's thin mantle also affects convection patterns. Geodynamical models suggest a short wavelength pattern with many small convection cells (Fig. \ref{Fig:Convection}a). Another aspect influencing the convective pattern are surface temperature variations caused by Mercury's unique orbit, which leads to uneven insolation \citep[][Chapter 3 and references therein]{solomon2018mercury}. Mercury's 3:2 spin-orbit resonance results in cold poles at the geographic poles and hot poles at the equator (centered at 0° and 180°). Given Mercury's thin mantle, this temperature distribution propagates through the mantle down to the core-mantle boundary and mildly affects the convection pattern \citep[][Chapter 19 and references therein]{solomon2018mercury}. As a result convection is characterized by smaller, colder plumes beneath colder surface areas and warmer upwelling plumes beneath regions with high surface temperatures (Fig. \ref{Fig:Convection}a).

\subsection*{Venus}
Often called the Earth's twin due to its similar mass, radius, and potential composition, Venus is a volcanic world with a young surface dominated by volcanic features. Although its geodynamic regime is not well understood, models indicate that magmatism has significantly influenced its evolution. Earlier studies suggested a global catastrophic event renewed the surface, while recent models propose gradual surface recycling through magmatic intrusions (the plutonic squishy lid regime). These intrusions create strong temperature variations in the crust and lithosphere (Fig. \ref{Fig:Convection}b), affecting viscosity and promoting surface recycling on smaller scales than Earth's plate tectonics \citep{Lourencco2023}. The intense magmatic activity and small-scale surface recycling characteristic for the plutonic squishy lid regime show a good agreement with the widespread volcanic features and tectonic motion observed on Venus.

The deep interior of Venus remains poorly understood. Studies indicate a higher correlation of gravity and topography for long wavelengths and a globally large apparent depth of compensation of topographic loads when compared to Earth, indicating support by mantle flow \citep[e.g.,][]{rolf2022}. This information can be used to constrain the mantle's viscosity structure. Models fitting the observed data propose a low viscosity layer beneath the thermal lithosphere, possibly linked to partial melting \citep[][and references therein]{Gillmann2024}. Such a mechanically weak layer is also present on the Earth (i.e., the asthenosphere), suggesting that Venus and the Earth are more similar than previously thought. Although solid-solid phase transitions are expected to occur, a sudden viscosity increase at the ringwoodite-bridgmanite transition is highly debated on the Earth and inconsistent with the analysis of gravity and topography data on Venus.

On Venus, high surface temperatures influence the tectonic regime and heat transport from the deep interior, making atmosphere-interior coupling more complex than on other terrestrial planets. Whether Venus always had such high temperatures or once had milder conditions that could support liquid water is debated. Some models suggest liquid water was possible for billions of years, with today's dense atmosphere and extreme temperatures resulting from a runaway greenhouse effect later in its history. The interior dynamics are believed to play a crucial role in atmospheric evolution, as melt from the mantle transports gases like water and carbon dioxide to the surface, where they escape into the atmosphere \citep[][and references therein]{Gillmann2024}.

\subsection*{The Moon}
Apollo seismic measurements revealed that the Moon possesses a small core, with the smallest core-to-planet radius ratio among terrestrial planets. Analysis of lunar samples indicated that the Moon was initially molten with a deep global magma ocean that shaped its interior \citep[][Chapter 3, and references therein]{Neal2024}. The crystallization of this magma ocean led to an unstable density stratification in the mantle that is prone to overturn. While earlier studies treated magma ocean crystallization and mantle overturn as separate events, recent models suggest that mantle overturn began during crystallization, indicating more efficient mixing than previously thought. This overturn is believed to have created geochemical reservoirs in the Moon's interior, which may explain some of the geochemical anomalies observed at the lunar surface and in lunar rocks \citep[][Chapter 4 and references therein]{Neal2024}.

The early history of the Moon significantly influenced its interior evolution. The asymmetry in crustal composition and volcanic activity between the far- and nearside is linked to the lunar mantle's evolution. Studies \citep[][and references therein]{Neal2024} suggest that a layer rich in heat-producing elements, a remnant of the lunar magma ocean solidification stage, may have remained beneath the PKT (Section \ref{sec:crust-planets}), an anomalous area on the lunar surface. Geodynamic models indicate that this hot near-surface layer results in a warmer nearside mantle (Fig. \ref{Fig:Convection}c) with concentrated magmatic activity and higher-than-average heat flows in the PKT region \citep[][Chapter 6, and references therein]{Neal2024}. Consequently, Apollo heat flow measurements in this area may not represent the Moon's average surface heat flow.

\subsection*{Mars}
The most striking feature of Mars is the difference in topography between the northern lowlands and southern highlands, which is thought to correlate with the crustal thickness -- thinner in the north and thicker in the south (Section \ref{sec:crust-planets}). The thicker southern crust acts as a thermal blanket, leading to a warmer interior and thinner lithosphere compared to the north (Fig. \ref{Fig:Convection}d). While the timing of its formation is still debated, the martian crustal dichotomy is thought to be one of the oldest features on the planet \citep[][and references therein]{Grott2013}. Several formation scenarios have been proposed, including a degree-one mantle convection pattern, i.e., a pattern characterized by one large plume \citep[][and references therein]{Grott2013}. However, seismic data from InSight suggests a thinner mantle and a core-to-planet radius ratio similar to Earth's (about 0.543 for Mars vs. 0.547 for Earth), making it unlikely for a degree-one convection pattern to naturally from, as such a pattern is more prone to develop in a Moon-like interior structure, where the core-to-planet radius ratio is small (i.e., ~0.201). While so far, the formation mechanism of the martian dichotomy is still debated, perhaps the most compelling scenario involves a large impact event followed by a subsequent more exotic mantle convection pattern.

Mars seismic data has been used to place constraints on the temperature of the mantle, as the latter ultimately affects the propagation time of seismic waves \citep{Huang2022}. While a range of estimates are available, mostly indicating a colder mantle than previously thought, care must be taken with these interpretations, as trade-offs between temperature and mantle chemical composition exist. In fact, Mars’ interior seems to be characterized by strong temperature variations. A cold interior is required to produce a large elastic lithosphere thickness necessary to explain the absence of surface deflection beneath the load induced by the north polar ice cap \citep[][and reference therein]{Grott2013}. On the other hand, geological and geophysical data indicate that volcanism has been recently active \citep[within the last few million years,][and reference therein]{Grott2013}, suggesting that melt may still be produced in the interior today, requiring high temperatures at least locally.

\section{Cores}\label{chap:cores} 
\begin{figure}[h!]
\centering
\includegraphics[width=\textwidth]{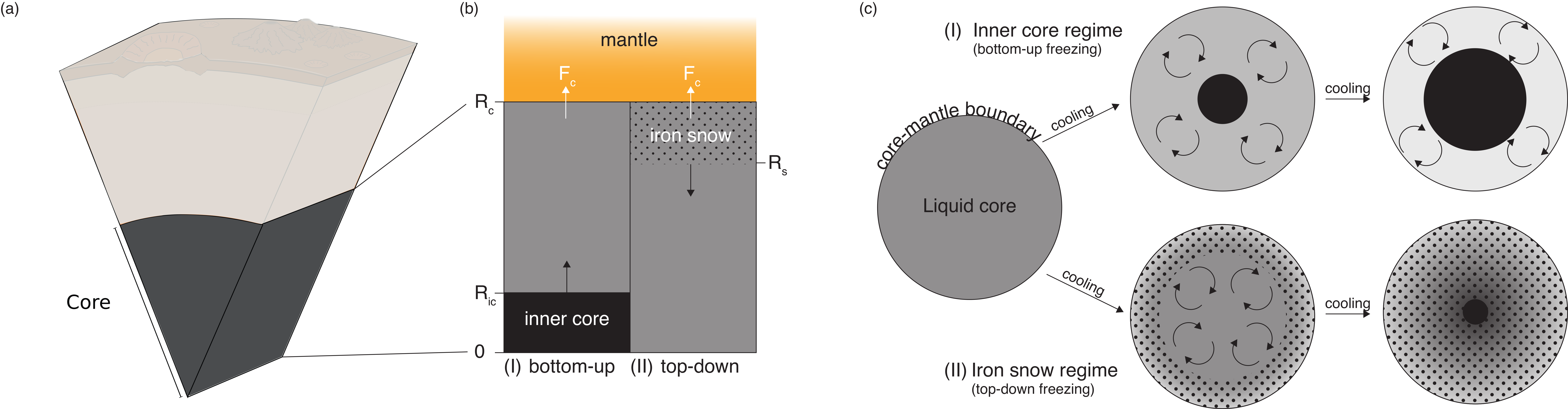}
\caption{Schematic view of the metallic core layer. (a) Location of the core layer with respect to other major layers. (b) Bottom-up (inner core growth regime) and top-down (iron snow regime) core freezing scenarios of a core cooling with heat flux $F_c$ with solid-liquid boundaries: $R_i$ inner core boundary and $R_s$ lower boundary of the the iron snow zone. (c) Sketch of the temporal evolution of inner core growth and iron snow. Black areas indicate solid iron. Gray areas depict liquid core regions. Curved arrows show chemical convection. In the snow zone (black dotted area) a stable chemical gradient develops.}
\label{Fig:core-overview}
\end{figure}

Metallic cores constitute the deep interior of terrestrial planets (Fig.~\ref{Fig:core-overview}a). They form -- if temperatures are high enough -- during the differentiation of an initially homogeneous planet as the heavy metals, mostly iron (Fe) and nickel (Ni), separate from the lighter silicate materials. After formation, the core typically loses heat across the core-mantle boundary, the amount of which depends on the efficiency of heat transport in the overlying mantle (Section \ref{sec:mantle_planets}). The loss of heat can trigger various processes in the core such as convection and crystallization. These processes can lead to a core dynamo and the generation of a global magnetic field. In the following, we will briefly discuss the constituents of terrestrial cores (Section \ref{sec:core_composition}), the general concepts of core cooling (Section~\ref{sec:core_cooling}) as well as the processes leading to a core dynamo (Section \ref{sec:core_dynamo}). Finally, in Section \ref{sec:core_planets} we compare the state of knowledge for the cores of the terrestrial planets.

\subsection{Core composition}\label{sec:core_composition}
The composition of the core constrains material properties or partitioning behavior during core freezing, which are crucial for accessing the core cooling history. From seismic measurements, it is known that the metallic core of Earth contains mostly heavy elements (iron or nickel) and additional light alloying elements. However, it is not known what exactly these light alloying elements are and what their concentration is. Potential candidates are sulfur, carbon, silicon, oxygen, hydrogen, sodium or phosphorus~\citep{hirose2013composition}. Additionally, the existence of small percentages of radioactive elements (e.g., potassium) has been discussed based on experimental studies~\citep{murthy2003experimental}. Even though seismic measurements at such level of detail are not available for other terrestrial planets or moons, it is reasonable to assume that their cores are also a mixture of mostly heavy and some light elements. A better knowledge of the core composition requires planet formation models, experiments on element partitioning at high pressures and temperatures relevant for the planet's or moon's interior, and observations of planetary building blocks such as asteroids and comets.

Next to temperature and pressure, core composition (amount of iron/nickel and light alloying elements) defines the phase diagram and important material properties such as the density and the thermal conductivity. While the thermal conductivity controls the cooling behavior of the core, the phase diagram provides crucial information on stable phases at given temperature and pressure conditions. A common assumption for terrestrial core alloys is a mixture of iron and sulfur due to its simplicity compared to more complex systems. Sulfur is siderophile and abundant in the Solar System. The iron-sulfur alloy forms a binary eutectic alloy, which is characterized by an eutectic concentration and temperature. The latter defines the lowest melting temperature of the alloy. Whether iron (less sulfur than the eutectic concentration) or iron sulfide (more sulfur than the eutectic concentration) crystallizes depends on the amount of sulfur in the core. The nature of the crystallizing component can have crucial implications on the evolution of the core and consequently the generation of a global magnetic field (Section~\ref{sec:core_dynamo}). Thus, a good knowledge of the core composition is vital but unfortunately in most cases not existent.

\subsection{Core Cooling}\label{sec:core_cooling}
While the early cooling behavior of planetary cores is closely linked to the heat stored in the core right after core formation (separation of metals and silicates), the long-term cooling behavior is dictated by the evolution of the overlying mantle (Section \ref{sec:convection}). During core formation, heat dissipated due to viscous friction is stored in the metal-rich cores as well as the silicate mantle. The distribution of heat between metals and silicates depends on the exact materials and the process of material separation. If a lot of dissipated heat is stored in the metal, the core can be substantially hotter than the overlying mantle. In this case, the early core evolution is characterized by rapid cooling, i.e., a high heat flux from the core into the mantle. The other extreme is no temperature contrast between the core and the mantle, in which case, already the early cooling behavior of the core depends on how efficiently the overlying mantle cools. In general, an efficiently cooling mantle leads to an efficiently cooling core whereas a slowly cooling mantle leads to a slowly cooling core. One caveat to this is the heat is being produced in the core. This could be either due to radioactive elements in the core (radioactive decay) or crystallization of core materials (latent heat). In this case, efficient cooling of the mantle must not always lead to efficient cooling of the core as the excess (radioactive or latent) heat has to be removed from the core. 

Cooling of the core can have multiple consequences. The two most important are convection (large-scale movement of core material) and core freezing. There are different types of natural convection: (I) thermally and (II) chemically driven convection. In case of \textit{thermal convection}, temperature differences lead to density differences that in turn lead to gravitationally unstable situations resulting in large-scale movement of material, i.e. convection. In case of chemical convection, density differences are caused by different materials or different concentrations of the same material. Thermal convection in planetary cores requires a critical heat flux from the core into the mantle that is larger than the heat than can be theoretically conducted along the core's adiabatic gradient. An important parameter for constraining this critical heat flux is the thermal conductivity of the core material, because it defines the efficiency of heat conduction. The lower the thermal conductivity the less heat can be transported by conduction and vice versa. Thus, ultimately the likelihood of thermal convection in the core is determined by how efficient the core cools via conduction and how much heat is removed from the core by the mantle. If for example the critical heat flux is low and the mantle cools rapidly, the core will likely thermally convect. 

\textit{Chemical convection} in the core is usually a result of core freezing, i.e., the crystallization of core material at some location within the core. Note, that the core will not freeze at once due to a considerable pressure gradient and a composition involving more than one component. If during crystallization light elements are unequally distributed between the solid and the liquid phase, the residual liquid (residual of the freezing process) will be lighter or heavier than the surrounding liquids. The location of first-time freezing can lead to different freezing regimes (see Fig.~\ref{Fig:core-overview}b). If, for example, freezing occurs first at the center of the core -- as it is the case for Earth -- and the residual liquid is lighter than its surroundings, it will rise towards the top triggering chemical convection in the outer core. Chemical convection can theoretically continue until the entire core is frozen. On the other hand, if freezing occurs at the top of the core, solid iron crystals may form (iron snow zone) and sink towards the bottom of the core. Deeper in the core, where the local temperature is above the melting temperature, the crystals remelt and create an iron-rich liquid. This iron-rich liquid is heavier than the liquid of the deeper core and will ultimately lead to chemical convection in the deeper core. Chemical convection in the deeper core continues until the iron snow zone is present in the entire core. Once the snow zone extends across the entire core chemical convection ceases and a solid inner core starts to grow from the accumulating crystals (see Fig.~\ref{Fig:core-overview}c). A review of different crystallization scenarios is given in~\citet{breuer2015iron}. For the sake of completeness it should be noted that the release or removal of latent heat during crystallization can add a thermal component to the (chemical) convection caused by freezing.

\subsection{Core dynamo}\label{sec:core_dynamo}
Since the middle of the 20th century it is well established that complex fluid motions in planetary cores can lead to \textit{self-sustained dynamos}. At the very least, a dynamo requires the motion of electrically conductive fluid to convert kinetic to magnetic energy. Planetary cores provide electrically conducting material (iron/nickel), if they are (partially) liquid. In addition to that, convection as described in Section~\ref{sec:core_cooling}, is a source of kinetic energy. However, even if both prerequisites are fulfilled, a dynamo does not necessarily occur. Magnetohydrodynamic simulations have proven successful in replicating planetary dynamos. A result of these simulations are scaling laws that relate important parameters such as magnetic field strength and characteristic velocity of the flow to parameters controlling convective dynamos. Indeed, these simulations find that the convective power is the important parameter constraining the magnitude of the magnetic field and of the characteristic velocity. 

Another important finding is that the magnetic Reynolds number, which is directly proportional to the velocity of the flow and relates magnetic advection to magnetic diffusion, has to be above a critical value of $\sim$50 in order for a dynamo to develop. Hence, a dynamo requires the velocity of the flow to be large enough to overcome ohmic dissipation and thus to allow for magnetic induction to occur. Calculating the magnetic field strength as well as the magnetic Reynolds number requires an estimate of the convective power, which can be obtained through one-dimensional core evolution models. Finally, it should be noted that magnetohydrodynamic simulations apply much too high core viscosities and diffusivities due to the demanding computation power necessary for realistic values. Nevertheless, their scalings work suprisingly well and thus they are a crucial tool in understanding planetary dynamos.

\subsection{The Planets}\label{sec:core_planets}
\begin{figure}[h!]
\centering
\includegraphics[width=\textwidth]{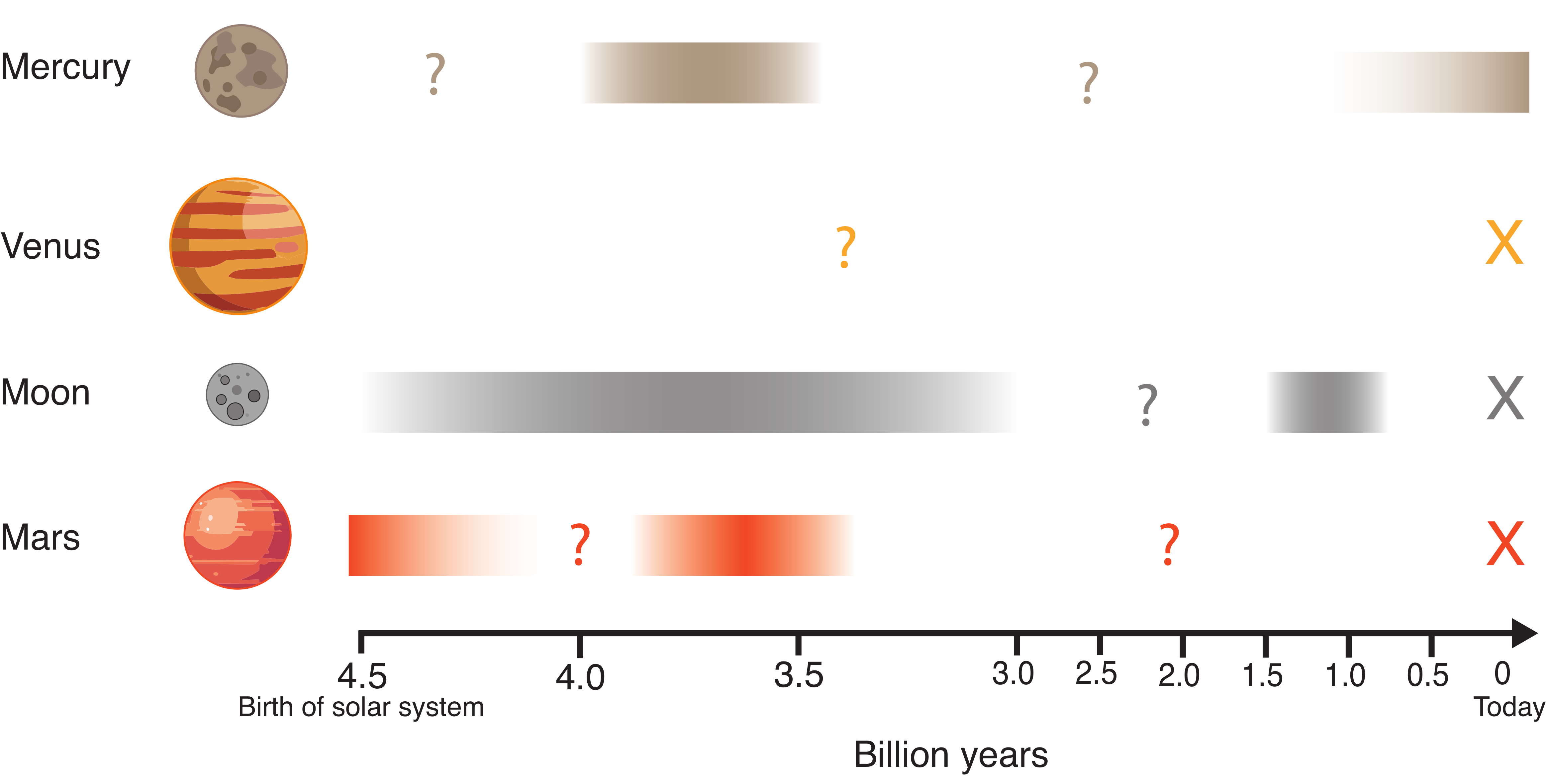}
\caption{Summary of the magnetic histories of Mecury, Venus, the Moon, and Mars. Question marks indicate a lack of data. The absence of a present-day magnetic field is illustrated by a cross at time 0. The colored bars represent the measured crustal magnetizations for the respective time periods, with the transparent ends indicating uncertainties in time.}
\label{Fig:core-magneticfield}
\end{figure}

The interior evolution and thus the evolution of the core of each terrestrial planet is unique. This results in very different magnetic histories summarized in Fig.~\ref{Fig:core-magneticfield}.

\subsubsection{Mercury}
Mercury shows signs of an internally generated present-day magnetic field as initially measured by the Mariner 10 mission and later confirmed by the MESSENGER mission \citep[][Chapter 1]{solomon2018mercury}. The observations show a peculiar small dipole moment for Mercury, which is a factor 10,000 smaller than that of Earth. Magnetohydrodynamic studies showed that the small dipole moment may be the result of a thermally stably stratified layer at the top of Mercury's core, which efficiently filters out the dipole components of the field \citep{christensen2006deep}. MESSENGER also found evidence of remanent crustal magnetization, for which the cratering record suggests a presence of the magnetic field from approximately 3.9 to 3.7\,Gyr ago \citep[][Chapter 19]{solomon2018mercury}. Thermal evolution models of Mercury suggest that the magnetic field was initially generated by thermal convection and later on until today by chemical convection due to the crystallization of an inner core similar to Earth's core. Since Mercury's core takes up almost 80\% of the planet's radius, core crystallization might be much more complex as freezing can occur simultaneously at multiple locations within the core~\citep{breuer2015iron}. Magnetic field measurements of the upcoming BepiColombo mission will yield much more detail on the field's morphology, allowing for more sophisticated thermal evolution and magnetohydrodynamic models.

\subsubsection{Venus}
So far, missions to Venus could neither detect a present-day magnetic field nor any crustal magnetization. This non-detection could be due to weathering of the crust occurring after a putative dynamo that can erase a previously acquired crustal remanent magnetization. However, weathering is thought to affect only shallow depths ($<$100\,m) and deeper crustal layers may still preserve magnetic signatures that could be detected by low orbit measurements from e.g., aerial platforms \citep{o2019detectability}. The lack of a present-day magnetic field could have various reasons: A too-slowly cooling core that might convect but not efficiently enough to allow for a dynamo or a primordial stratification of the core or an entirely solid core if initial temperatures were low \citep{o2018prospects}. Besides the unknown initial conditions, a major uncertainty is the regime of mantle convection, which ultimately defines the heat lost by the core (Section \ref{sec:mantle_planets}). In order to answer the question of an ancient dynamo for Venus, future work will have to focus on understanding exactly that cooling behavior. Upcoming missions like VERITAS and EnVision will help to constrain the geodynamic and thermal state of Venus' interior through determining the level of volcanic and tectonic activity. 

\subsubsection{The Moon}
The Moon has no present-day magnetic field. The Apollo missions, however, brought back many samples, which have been extensively analyzed for paleomagnetic signatures~\citep{Neal2024}, Chapter 5. The paleomagnetic record shows potential magnetic activity throughout a large part of lunar history featuring a peculiar high intensity field epoch (comparable to that of Earth today) at the very beginning, which transitions into a low field epoch around 3.56 to 3.19 Gyr ago. The most recent evidence for magnetization has been found to have occurred 0.8 Gyr ago. Thermal core evolution models that consider thermal and chemical convection are challenged in two ways: (1) These models are not able to explain the high field intensities for reasonable lunar core sizes. (2) The presence of the lunar dynamo until very recently may be hard to reconcile with the power available in the interior. Dynamos driven by mechanical forces such as e.g., precession or libration have been discussed in order to explain the high-field intensities, as such dynamos can produce a much larger field strength. The prolonged presence of the lunar dynamo might be explained by a start-stop mechanism, which involves the repeated onset and cessation of convection in the core. Ultimately, the lunar magnetic history can probably best be explained by a combination of different mechanisms, whose exact interplay is the subject of current research.    

\subsubsection{Mars}
Mars does not feature an internally generated magnetic field today, but it does show a magnetized crust, which hints towards an active field in the past \citep{10.3389/fspas.2022.895362}. Data acquired by magnetometers orbiting Mars indicate that the crustal magnetization reflects the martian dichotomy, with a stronger field on the southern hemisphere. Large impact basins such as Hellas, Argyre, Utopia, and Isidis show only weak or no magnetization, which suggests a lack of dynamo at the time of their formation ($\sim$4\,Gyr ago). Recent data indicates an active magnetic field up to at least 3.7\,Gyr ago \citep[][and references therein]{10.3389/fspas.2022.895362}, posing important constraints for the cooling of the martian interior. The absence of an inner core, which is supported by seismic data from the InSight mission (no detection of an inner core) suggests that the once-active dynamo on Mars was driven by thermal convection in the liquid core. With time, the early field then ceased due to inefficient mantle cooling as supported by mantle evolution models, which yield temperatures at the core-mantle boundary above the melting temperature~\citep{10.3389/fspas.2022.895362}. However, the most recent data on core size and composition from the InSight mission motivates future studies on core evolution to refine timing estimates and core crystallization scenarios. 

\section{Conclusions}\label{chap:conclusions} 
Planetary surfaces and interiors are intimately linked and understanding their complex interplay requires a multi-disciplinary approach involving fields such as geology, geophysics, geodynamics, and geochemistry. 
By combining mission data with numerical models and laboratory experiments, our knowledge about the surface and interior of rocky planets has significantly increased, allowing us to answer fundamental questions about the formation and evolution of planets but at the same time opening new questions whose answers necessitate new missions and theories. Some of these questions, such as why does Mercury possesses an active magnetic field today?, what is the level of volcanic and tectonic activity on Venus?, how much water is there on the Moon?, what do samples returned from Mars tell us about its geological history and astrobiological potential?, will be addressed by analyzing and interpreting data to be returned by missions planned or already flying to their targets. However, more missions beyond the currently planned ones are needed to unlock the mysteries hidden deep in the interior of planets. For example, high resolution gravity field missions such as GRAIL on the Moon and GRACE on the Earth have shown their potential to reveal the subsurface and interior of planets with unprecedented detail. Seismic sensors and seismic networks could greatly improve our understanding of the deep interior of rocky planets that is otherwise impossible to directly access. Magnetic field measurements at or close to the surfaces of terrestrial planets could reveal the traces of past magnetic fields, too weak to be detected by orbiters, while transient electromagnetic sounding could inform about the distribution of subsurface water. Understanding the surfaces and interiors of our planetary neighbors can teach us about the diversity of evolutionary paths that a terrestrial planet might follow and help us to put our own planet in a global context of planetary evolution.

\bibliographystyle{OtherDocs/Harvard}
\bibliography{reference}

\end{document}